\documentclass[prb,floatfix,twocolumn,showpacs,aps]{revtex4}
\usepackage{graphicx}
\usepackage{dcolumn}
\usepackage{amsmath}
\usepackage{natbib}
\newcommand{\DT} {$\Theta_{\textrm{D}}$}
\begin{document}
\title{Studies of Vibrational Properties in Ga Stabilized $\delta$-Pu \\ by Extended X-ray Absorption 
Fine Structure}
\author{P. G. Allen}
\email{allen42@llnl.gov}
\affiliation{Seaborg Institute for Transactinium Science, Lawrence Livermore National Laboratory, 
P.O. Box 808, Livermore, California 94551}
\author{A. L. Henderson}
\affiliation{Seaborg Institute for Transactinium Science, Lawrence Livermore National Laboratory, 
P.O. Box 808, Livermore, California 94551}
\author{E. R. Sylwester}
\affiliation{Seaborg Institute for Transactinium Science, Lawrence Livermore National Laboratory, 
P.O. Box 808, Livermore, California 94551}
\author{P. E. A. Turchi}
\affiliation{Materials Science and Technology Division, Lawrence Livermore National Laboratory, 
P.O. Box 808, Livermore, California 94551}
\author{T. H. Shen}
\affiliation{Materials Science and Technology Division, Lawrence Livermore National Laboratory, 
P.O. Box 808, Livermore, California 94551}
\author{G. F. Gallegos}
\affiliation{Materials Science and Technology Division, Lawrence Livermore National Laboratory, 
P.O. Box 808, Livermore, California 94551}
\author{C. H. Booth}
\affiliation{Chemical Sciences Division, Lawrence Berkeley National Laboratory, 
Berkeley, California 94720}

\date{Phys. Rev. B in press as of April 15, 2002}

\begin{abstract}
Temperature dependent extended x-ray absorption fine structure (EXAFS) 
spectra were measured for a 3.3 at\% Ga stabilized Pu alloy over the 
range $T$= 20 - 300 K. EXAFS data were acquired at both the Ga $K$-edge and the Pu 
$L_{\textrm{III}}$-edge.   Curve-fits were performed to the first shell interactions to 
obtain pair-distance distribution widths, $\sigma$, as a function of 
temperature.  The temperature dependence of $\sigma(T)$ was 
accurately modeled using a correlated-Debye model for the 
lattice vibrational properties, suggesting Debye-like behavior in 
this material.  Using this formalism, we obtain pair-specific 
correlated-Debye temperatures, $\Theta_{\textrm{cD}}$, of 110.7 
$\pm$1.7 K and 202.6 $\pm$3.7 K, for the Pu-Pu and 
Ga-Pu pairs, respectively.  The result for the 
Pu-$\Theta_{\textrm{cD}}$ value compares well with 
previous vibrational studies on $\delta$-Pu.  In addition, our results represent 
the first unambiguous determination of Ga-specific vibrational properties in PuGa alloys, 
i.e, $\Theta_{\textrm{cD}}$ for the Ga-Pu pair.  Because 
the Debye temperature can be related 
to a measure of the lattice stiffness, these results indicate the Ga-Pu 
bonds are significantly stronger than the Pu-Pu bonds.  This effect has 
important implications for lattice stabilization 
mechanisms in these alloys.
\end{abstract}
\maketitle
\section{Introduction}
Plutonium in its elemental form presents a complicated picture 
of phase stability.\cite{Miner90}  Indeed, Pu may adopt one of 
six crystallographically different phases ($\alpha$ ,  $\beta$,  $\gamma$ ,  
$\delta$ ,  $\delta'$,  and $\epsilon$) upon heating from room
temperature to its melting point of 913 K.   
This complex phase behavior has been explained, in part, 
by the behavior of the Pu 5$f$ orbitals which fluctuate between itinerant 
and localized behavior.  In this heuristic framework, 
elemental Pu represents the transition point along the actinide series 
from the delocalized electronic nature of the early actinides (Ac-Np) to localized, 
lanthanide-like $f$-orbital character observed for the heavier 
actinides.\cite{Smith83}
Theoretical calculations currently support the hypothesis that the 
5$f$ orbitals are delocalized for elements lighter than Pu in the actinide 
series and localized for those heavier than Pu.\cite{Willis92}  This view of the 
relationship between electronic and crystallographic structure is 
reinforced by comparing the complex phase behavior and structures observed 
for pure U, Np, and Pu metals as opposed to the relatively simple structures 
observed for the elements Am and beyond. \

The structural and electronic relationships that exist between the $\alpha$ 
and $\delta$ phases have received much attention due to some unusual observations.   
Pure face-centered cubic (fcc) $\delta$-Pu is stable from 593 to 736 K, and 
exhibits a 25\% increase in volume relative to that of the ground state 
phase,\cite{Miner90} monoclinic $\alpha$-Pu.  However, it is well known that 
the fcc structure can be stabilized down to ambient temperature by the 
addition of small amounts ($\sim$3-9 at\% ) of alloying elements such as Al, Ga, 
In, Sc, and Ce.\cite{Chiotti81}  However at lower impurity atom 
concentrations, the $\delta$ phase converts 
directly to the $\alpha$ form upon cooling, possibly through a martensitic phase 
transformation.\cite{Adler92,Zocco90} \

Other than a tendency to form trivalent cations, relatively little is 
known about the mechanism of these so-called 
``$\delta$-stabilizers''.\cite{Gschneidener61,Adler91}  Band 
structure studies have suggested that Al, Ga and Sc impurities diffuse the 
5$f$ bands, thereby removing their involvement in bonding and leading to the 
stability of a more common, $d$-bonded transition metal-like phase.\cite{Weinberger85}  
Becker {\em{et al.}} have employed an ab initio LDA approach to study the lattice 
relaxation in a Pu$_{31}$Ga supercell cluster, and find evidence of a lattice 
contraction around the Ga site which relaxes the bonding constraints for 
the neighboring Pu atoms.\cite{Becker98}   Other theoretical 
works\cite{Turchi99} point to a 
substantial level of 5$f$ localization in the $\delta$-phase\cite{Penicaud97} 
and have speculated on a Kondo-like model for the involvement of the 5$f$ 
electrons.\cite{Fournier96} \ 

Analysis of the vibrational properties in these materials is also 
important for understanding $\delta$-phase stablization. 
Unfortunately, measurement of the phonon dispersion using conventional 
inelastic neutron scattering techniques is troublesome given the 
difficulty in growing high-quality single crystals.  An alternative approach is 
to evaluate the Debye temperature, \DT, using various structural 
techniques.   Ledbetter {\em{et al.}}\cite{Ledbetter76} employed ultrasonic wave measurements 
of elastic constants for 3.3 at\% Ga $\delta$-Pu and calculated a 
\DT value of 115 K.  The thermal behavior of 5.0 at\% Al $\delta$-Pu 
was studied by temperature-dependent neutron powder 
diffraction\cite{Lawson94} yielding a similarly low value of \DT=132 K.  
More recently, Lynn {\em{et al.}}\cite{Lynn98} studied a 3.6 at\% Ga $\delta$-Pu 
sample using neutron-resonance Doppler spectroscopy, a technique that can 
determine element specific values for \DT.  The experiments determined a 
Pu-specific value of \DT=127 K, and assigned a Ga specific value of 
\DT=255 K, although with relatively large errors ($\pm$22 K).  As a 
test of the Debye model, one can check the assumption of equal force 
constants for the Pu and Ga sites by comparing the Ga-\DT~ to the $\sqrt{m_{\textrm{Pu}}/m_{\textrm{Ga}}}$ 
weighted value of 236 K derived from the Pu-\DT value.  Thus, this 
comparison does not exclude the possibility that the Ga atoms experience 
a stiffer force field compared with the Pu atoms. \

In this article, we present temperature dependent EXAFS (extended x-ray 
absorption fine-structure) spectroscopic results for 3.3 at\% Ga $\delta$-Pu as a 
means of discerning differences in the vibrational character of the Ga and 
Pu sites.   Previously, EXAFS studies on Ga stabilized 
$\delta$-Pu have focused on isothermal measurements at either the Ga $K$- or the 
Pu $L_{\textrm{III}}$-edges, and have revealed some important 
effects.\cite{Cox95, Faure96, Richard98}  In 
general, these studies indicate that the Ga atoms reside in their expected 
fcc lattice positions although there is an appreciable lattice contraction 
observed for the Ga-Pu bonds ($\sim$3-4\%).  Surprisingly, the contraction in 
Ga-Pu bonds is significantly larger than the collapse calculated 
theoretically\cite{Becker98} or the contraction expected from Vegard's law based on 
a simple substitutional alloy.  In addition, EXAFS data show increased 
disorder for the Pu-Pu near neighbor interactions.  These results indicate 
that there are unexpected, site-specific lattice effects occurring in 
these materials.  \

The outline of the paper is as follows: details of sample preparation 
and EXAFS experimental setup and data analysis are discussed in Sec. 
II.   The results of curve-fitting analysis and modeling the 
$\sigma(T)$ temperature dependence using the correlated Debye model 
are presented in Sec. III.  A discussion of the results and their 
relation to other \DT~studies is presented in Sec. IV, and the conclusions are given in Sec. V.

\section{Experimental Details}
\subsection{Sample Preparation}
A $\sim$6 $\mu$m thick $^{239}$Pu foil (3.3 at \% Ga) was prepared from 20 year old 
material by melting to remove accumulated helium, followed by subsequent 
annealing, cutting, and rolling.  The foil was further homogenized at 
450 $^{\circ}$C for $\sim$100 hrs to ensure that single-phase, $\delta$-Pu was produced.  
Transmission x-ray diffraction was performed at LLNL and confirmed the presence of the fcc 
phase, with no significant amounts of other Pu phases present.  In 
preparation for EXAFS analysis, the foil was electropolished to remove any 
accumulated oxide material on the surface.  The sample was then 
encapsulated under argon using a specially designed, triple containment 
x-ray compatible cell manufactured by Boyd Tech.  The first level of containment consisted of 
coating the sample in a thin film of liquid polyimide solution that was 
allowed to air dry directly on the sample.  The foil was then mounted onto 
an aluminum frame and sealed within two additional layers of x-ray 
transparent Kapton windows (0.010" thick).  The windows were clamp-mounted 
onto the aluminum body with stainless steel window frames and using indium 
wire as a vacuum seal material.  The 
triple-contained sample was subsequently mounted in an open cycle liquid 
helium flow-cryostat for variable temperature EXAFS measurements.    
Temperature measurement errors are within $\sim$1 K, and are stable 
within $\sim$0.2 K.

\subsection{EXAFS Data Acquisition and Analysis}

Plutonium $L_{\textrm{III}}$- and gallium $K$-edge x-ray absorption spectra were 
collected at the Stanford Synchrotron Radiation Laboratory (SSRL) on 
wiggler beamline 11-2 under normal ring operating conditions using a 
nitrogen-cooled Si (220), half-tuned, double-crystal monochromator 
operating in unfocussed mode.  The vertical slit height inside the x-ray 
hutch was 0.3 mm which reduces the effects of beam instabilities and 
monochromator glitches while providing ample photon flux.  The Pu 
$L_{\textrm{III}}$-edge spectra were measured in transmission mode using Ar-filled 
ionization chambers.  The Ga $K$-edge spectra were measured in 
fluorescence mode using a 30-element Ge array solid state detector developed 
by Canberra Industries.  The detector 
was operated at $\sim$75 kHz per channel, and the signals were digitally 
processed using the DXP 4C/4T developed by X-ray Instrumentation 
Associates.  \

XAFS raw data treatment, including calibration, normalization, and 
subsequent processing of the EXAFS and XANES (x-ray absorption 
near-edge structure) spectral regions was performed by standard methods 
reviewed elsewhere\cite{Hayes82,Li95b} using the EXAFSPAK suite of programs developed by 
G. George of SSRL.  Typically, three XAFS scans (transmission or 
fluorescence) were collected from each sample at each temperature and the 
results were averaged.  The spectra were energy calibrated by 
simultaneously measuring the absorption spectrum for the reference samples 
PuO$_{2}$ or Ga$_{2}$O$_{3}$.  The energies of the first inflection points for the 
reference sample absorption edges, $E_{r}$, were defined at 18053.1 
eV (Pu $L_{\textrm{III}}$) and 10368.2 eV (Ga $K$).  The EXAFS threshold 
energies, $E_{0}$, were defined as 18070 eV and 10385 eV for the Pu 
and Ga edges, respectively.  Nonlinear least-squares curve-fitting was performed on the 
$k^{3}$-weighted EXAFS data using the EXAFSPAK programs.  \

The EXAFS data were fit using theoretical phase and amplitude 
functions calculated from the program FEFF8.1 of Rehr {\em{et al.}}\cite{Rehr91a,Rehr91b}
All of the Pu-Pu interactions were modeled using single scattering (SS) 
paths derived from the model compound, unalloyed $\delta$-Pu, 
$a_{\textrm{fcc}}$=4.6371 \AA.\cite{Ellinger56}  The Ga-Pu SS interactions were 
modeled by using the same model compound structure and replacing the central 
absorbing atom with Ga.  An initial series of fits was done on 
the raw Pu $L_{\textrm{III}}$ and Ga $K$-edge $k^{3}$-weighted data sets using the expected 
fcc near-neighbor interactions at 3.28, 4.64, 5.68, and 6.56 \AA~and 
fixing the coordination numbers at 12, 6, 24, 12, and 24, respectively.
The results (especially at low T) confirmed the presence of 
the fcc structure and showed no evidence for other unusual structural 
effects.  That is, we observed no phase changes (i.e., $\delta\to\alpha$) or previously postulated 
metastable impurity phases.\cite{Conradson98, Conradson00}   This first level of analysis was 
also used to establish values for $S_0^2$ and $\Delta$$E_{0}$ by fixing coordination 
numbers, $N$, and allowing $S_0^2$, $\Delta$$E_{0}$, $\sigma^{2}$, and 
$R$ to vary.  The final values used were taken from averaging over the 
range of temperatures studied: for Pu $S_0^2$ = 0.55, 
$\Delta$$E_{0}$= -12 eV; and for Ga, $S_0^2$ = 0.85, $\Delta$$E_{0}$= -10 eV. \

As a result of the preliminary analyses, all of the subsequent fits 
described here focussed on isolating the behavior of the first shell 
Pu-Pu and Ga-Pu interactions in a highly constrained manner.  Thus 
fits were done on Fourier-filtered data using the same fixed $S_0^2$ and $\Delta$$E_{0}$ values for 
all temperatures, along with a fixed coordination number of $N$=12 for the first 
shell.  The ability to fix $\Delta$$E_{0}$, $N$, and $S_0^2$ helps to 
avoid correlation problems between the fit parameters and to establish 
more consistently any changes in 
$\sigma^{2}$ and $R$ that may occur as functions of temperature. \
\begin{figure}
\includegraphics[width=2.7in] {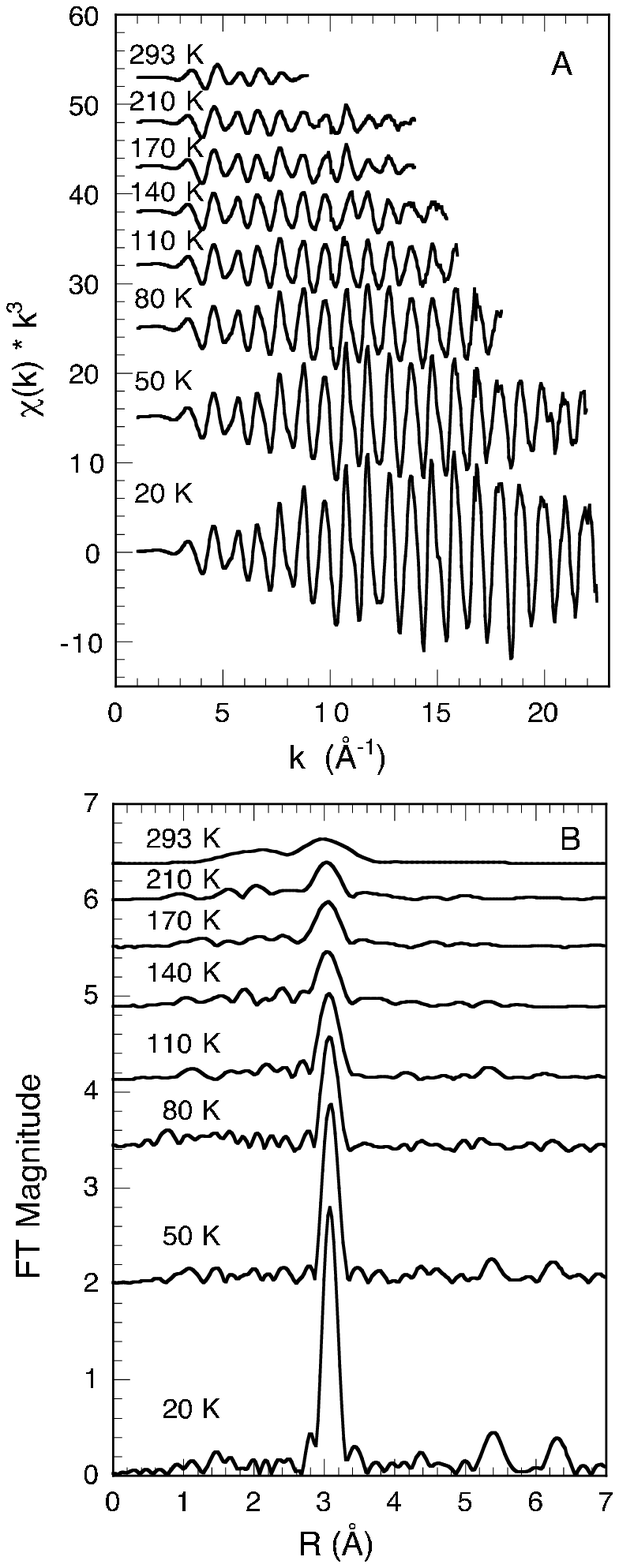}
\caption{Pu $L_{\textrm{III}}$-edge $k^{3}$-weighted EXAFS data 
(A) and the corresponding Fourier transforms (B) for the 
$\delta$-Pu sample as a function temperature.  Data 
were acquired in transmission mode and Fourier transformed over 
the ranges shown in the figure.}
\label{puXAFS}
\end{figure}

\section{Results}
\subsection{EXAFS Raw Data and Curve-Fitting}
Figures 1A and 1B show the raw $k^{3}$-weighted Pu EXAFS $L_{\textrm{III}}$ 
data and the corresponding Fourier transforms (FT) for the $\delta$-Pu 
sample as a function of temperature.   The FT represents 
a pseudo-radial distribution function and the peaks are shifted to lower 
$R$ values compared to real interaction values as a result of the phase shifts associated with the 
absorber-scatterer interactions ($\sim$0.1-0.2 \AA\ for Pu-Pu).   As the sample is cooled 
from 293 K to 20 K, the EXAFS scattering amplitude increases 
systematically due to decreased thermal motion of the atoms in the 
lattice.  As a result, the effective measurable $k$-range increases 
substantially at lower temperature.  This effect is equally visible 
in the corresponding FTs.  As the temperature is lowered, the intensity of the FT peaks 
increases dramatically, and the spectra reveal a pattern consistent with that expected 
for a fcc lattice.  The first shell Pu-Pu peak seen at $\sim$3.1 \AA\ 
corresponds to 12 Pu near neighbors in the fcc structure and increases 
in maximum peak height while also narrowing with decreasing temperature.  
The second, third, and fourth shell peaks at $\sim$4.4, 5.4 and 6.3 
\AA~which correspond to real interactions at 4.64, 5.68, and 6.56 \AA~are 
clearly affected by thermal effects and become distinguishable only 
below 80 K. \

\begin{figure}
\includegraphics[width=2.7in] {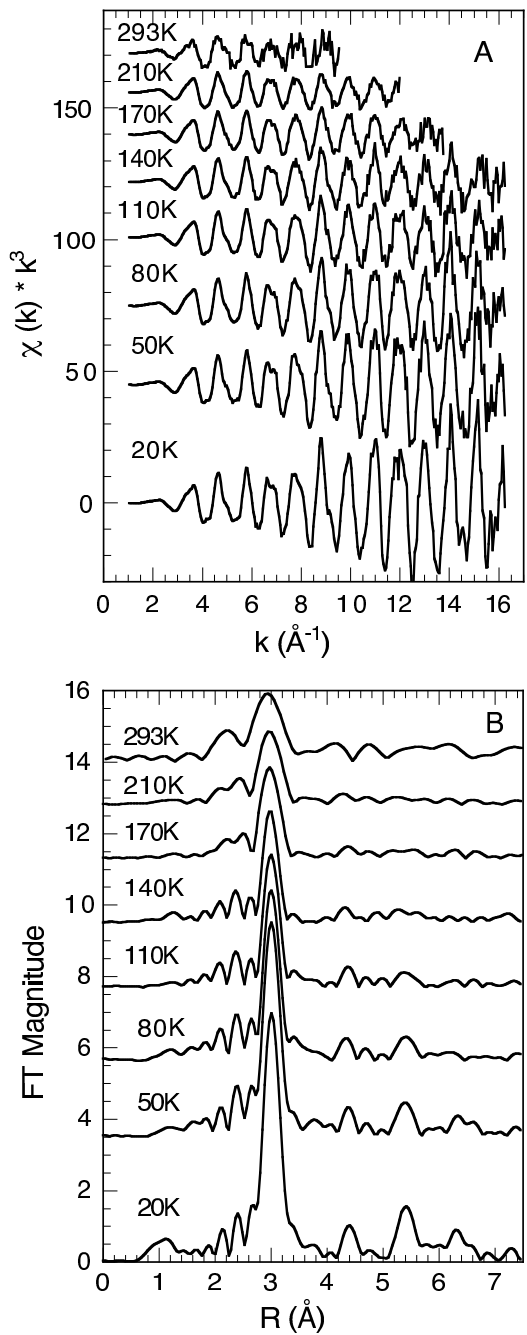}
\caption{Ga $K$-edge $k^{3}$-weighted EXAFS data 
(A) and the corresponding Fourier 
transforms (B) for the $\delta$-Pu sample as a function temperature.  
Data were acquired in fluorescence mode and Fourier transformed over 
the ranges shown in the figure.}
\label{gaXAFS}
\end{figure}

The Ga $K$-edge raw $k^{3}$-weighted EXAFS and corresponding FTs are shown in 
Figures 2A and 2B.  This series of spectra qualitatively exhibit the same 
thermal behavior observed in the Pu $L_{\textrm{III}}$-edge EXAFS 
data, that is, increased scattering amplitude along with detection of more distant neighbors are 
apparent with decreasing temperature.  The spectra are dominated by 
the peak at $\sim$3.0 \AA, attributed to the first shell Ga-Pu interactions 
($N$=12).  At lower temperatures, the FTs reveal second, third, and fourth shell Pu neighbor 
interactions at $\sim$4.4, 5.4, 6.3 \AA, respectively.  In contrast to 
the behavior observed in the Pu EXAFS, the detection of 
these longer range interactions is retained up to higher temperatures 
(ca. 110 K).  At this level of analysis, this observation suggests that the Ga-Pu interactions 
experience less thermal broadening than the corresponding Pu-Pu 
interactions, which is especially surprising given the decreased 
reduced mass of the Ga-Pu pair relative to the Pu-Pu pair.

\begin{table}
\caption{Pu $L_{\textrm{III}}$ and Ga $K$-edge EXAFS single shell curve fitting 
results.  The first shell interactions were isolated by Fourier 
transforming over the data $k$-ranges displayed in Figures 1A and 2A, 
and back-transforming over the range $R$=2-4 \AA.   Several 
parameters were held fixed as follows: for Pu and Ga, $N$=12; for Pu, 
$S_0^2$ = 0.55 and $\Delta$$E_{0}$= -12 eV; and for Ga, $S_0^2$ = 0.85 
and $\Delta$$E_{0}$= -10 eV.}
\begin{ruledtabular}
\begin{tabular}{lcccl}
    Sample & \multicolumn{2}{c}{Pu-Pu shell\footnote{Errors in $R$ 
    and $\sigma^{2}$ are estimated to be $\pm$0.005 \AA~ and 
    $\pm$10\% based on EXAFS fits to known model compounds, cf. Ref. 
    \onlinecite{Li95b}.}} & \multicolumn{2}{c}{Ga-Pu shell} \\
    Temp (K) &  {$R$ (\AA)}  & {$\sigma^{2}$ (\AA$^{2}$)} &  {$R$ (\AA)} &  {$\sigma^{2}$ (\AA$^{2}$)} \\\hline
20  &  3.290  &  0.00280  &  3.160  &  0.00275 \\
50  &  3.292  &  0.00381  &  3.160  &  0.00332 \\
80  &  3.293 &  0.00511  &  3.160  &  0.00395 \\
110  &  3.301  &  0.00680  &  3.160  &  0.00512 \\
140  &  3.301  &  0.00905  &  3.160  &  0.00609 \\
170  &  3.306  &  0.01039  &  3.163  &  0.00764 \\
210  &  3.316  &  0.01275  &  3.165 &  0.00953 \\
293  &  3.297  &  0.01821  &  3.171  &  0.01264 \\
\end{tabular}
\end{ruledtabular}
\label{fits}
\end{table}

Following initial inspection of the data (see Section II) we 
chose to focus our analyses on the isolated first shell contributions, 
in part, due to the high signal-to-noise obtained relative to the more 
distant shell interactions.  Thus by curve-fitting the Fourier-filtered first 
shell components from the Pu $L_{\textrm{III}}$ and Ga $K$ EXAFS, the 
local vibrational temperature dependence may be investigated.  The first 
shell interactions were isolated by Fourier transforming over the data 
$k$-ranges shown in Figures 1A and 2A, and back-transforming over the range 
$R$=2-4 \AA.  \

The curve-fitting results summarized in Table \ref{fits} serve to identify the different 
temperature dependent and static structural effects for the Pu and Ga sites in this 
material.  Figure 3 shows the resulting curve fits to the Fourier filtered Pu
and Ga EXAFS data measured at 20 K in $k$-space and  
$R$-space.  The strong coincidence between the data and the fits serves 
to illustrate the highly refined nature of the FEFF8.1 calculations as 
well as the appropriateness of the fitting procedure employed.  Because 
these fits were highly constrained, any changes 
in the data are assigned to changes in $R$ and $\sigma^{2}$.  The first 
effect revealed by these data is the overall contraction of Pu atoms 
around the Ga sites relative to the environment around the Pu sites.   This ``collapse''
of about 4\% is comparable to that observed by earlier EXAFS studies 
on $\delta$-Pu.\cite{Cox95, Faure96, Richard98}   The first shell bond 
lengths for each element appear to remain constant over the measured 
temperature range, at least within the quoted experimental error and 
under the constraints outlined above. \
\begin{figure}
\includegraphics[width=3.39in] {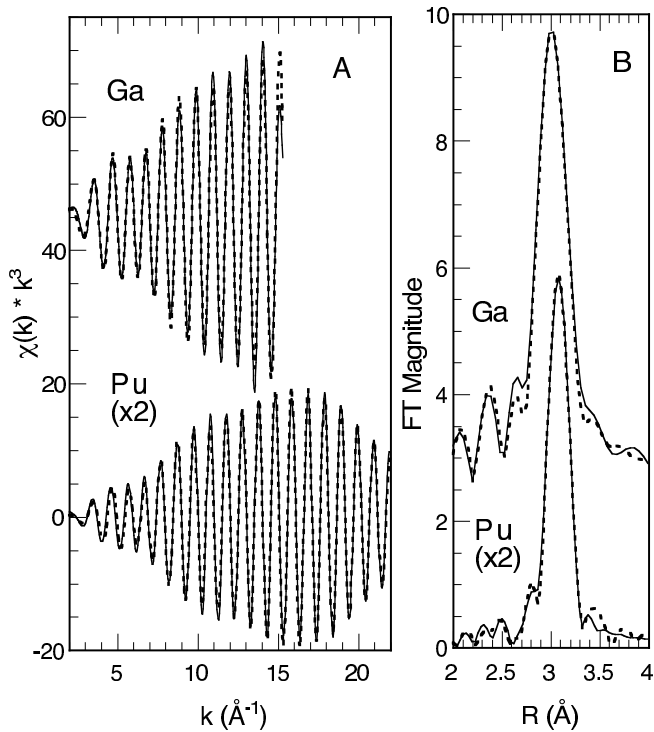}
\caption{Representative single shell curve fits (dotted lines) to the Pu $L_{\textrm{III}}$
and Ga $K$-edge data (solid lines) measured at 20K in (A) $k$-space and (B) 
$R$-space.  Data were Fourier-filtered from the raw spectra over the $k$ and $R$ ranges 
shown in the plots using Gaussian window functions with half-widths 
of 0.5 \AA$^{-1}$ and 0.05 \AA, respectively.  Pu data shown 
are multiplied by a factor of 2 for the purpose of comparison to the Ga data.}
\label{FITS}
\end{figure}

\subsection{Vibrational Analysis}
The most dramatic effect depicted in this data set is 
the difference in the temperature dependence between the Pu-Pu and Ga-Pu 
Debye-Waller factors, $\sigma^{2}$.  The values for $\sigma^{2}$ increase 
with temperature consistent with greater thermal disorder, as expected.
However, according to Table \ref{fits}, the Pu-Pu first shell shows 
much greater disorder at higher temperature than the corresponding 
Ga-Pu shell.  Moreover, this occurs in spite of the fact that the 
Ga-Pu pair has a lower reduced mass.  To study this effect more carefully, the temperature 
dependence of the Debye-Waller factors was modeled by employing the 
correlated-Debye Model to determine the Debye Temperature.\cite{Crozier88}
\begin{equation}
	\sigma_{meas}^{2}(T) = \sigma_{static}^{2} + F(T,
	\theta_{cD}).
\label{corr_debye}
\end{equation}
The temperature-dependent part of the Debye-Waller factor $F(T,
\theta_{cD}$) is given within the correlated-Debye model by
\begin{equation*}
F(T, \theta_{cD}) =\frac{\hbar}{2 \mu} \int \rho_{j}(\omega)
coth\left(\frac{\hbar \omega}{2 k_{B} T} \right)
\frac{d\omega}{\omega}
\end{equation*}
where $\mu$ is the reduced mass, $\theta_{cD}$ is the correlated Debye
temperature, and the phonon density of states at position $R_{j}$
is\cite{Beni76}
\begin{equation}
\rho_{j} = \frac{3 \omega^{2}}{\omega_{D}^{3}} \left[ 1 - 
\frac{sin(\omega R_{j}/c)}{\omega R_{j}/c}\right]
\label{dos}
\end{equation}
in which $\omega_{D}$ is the usual Debye frequency and $c=
\omega_{D}/k_{B}$ where $k_{B}$ is Boltzmann's constant.  The expression 
in brackets of Eq.  \ref{dos}
takes into account the correlated motion of the atom pairs.  

\begin{figure}
\includegraphics[width=3.0in] {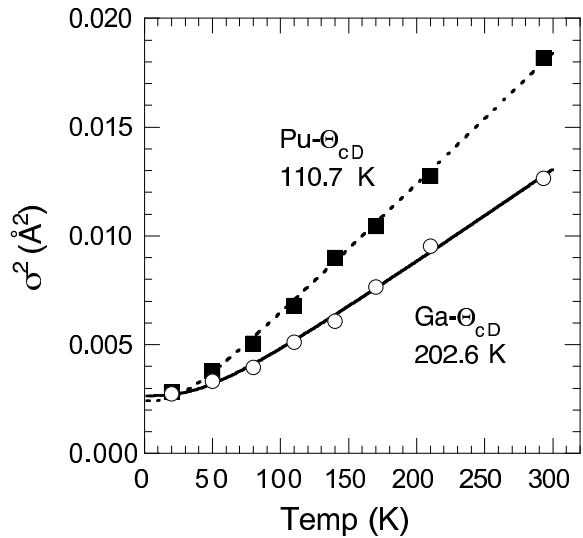}
\caption{Temperature dependence of EXAFS Debye-Waller factors for 
first shell Pu-Pu (upper data) and Ga-Pu (lower data) interactions 
plotted along with fits generated using the correlated Debye model.}
\label{CorrD}
\end{figure}

Using this formalism, we obtain the fits shown in Figure \ref{CorrD}, 
and determine pair-specific correlated-Debye 
Temperatures, $\Theta_{\textrm{cD}}$, of 110.7 $\pm$1.7 K and 
202.6 $\pm$3.7 K, for the Pu and Ga sites, respectively.  Estimates 
of the $\sigma_{static}^{2}$ static parameter are all consistent with no static 
disorder, and fall in the range of 0.0003 $\pm$0.0003 \AA$^{2}$.  The data 
appear to fit this model quite accurately, indicating that the 
local pair vibrations can be described using the correlated-Debye model.  However, the 
higher $\Theta_{\textrm{cD}}$ of the Ga-Pu pair with respect to the 
Pu-Pu pair may indicate the presence of a local lattice anomaly from a 
vibrational standpoint, which could be related 
to the observed lattice contraction measured around the Ga sites.
\section{Discussion}
It is instructive to compare the EXAFS $\Theta_{\textrm{cD}}$ results described here with 
those obtained from previous studies.  Table \ref{debye} summarizes the 
\DT~values for $\delta$-Pu determined from our study along with those 
from earlier works.  We also make special note of the differences in 
the methods used to determine these values, which is important 
for a proper interpretation.  The elastic values of \DT~are obtained by ultasound 
measurements of ``bulk'' lattice wave properties extrapolated down to 
T=0 K.\cite{Ledbetter76, Taylor68, Rosen69} The ``Bulk Modulus'' value uses a theoretical 
relationship\cite{Moruzzi88} where \($\DT$=41.63\sqrt{{r_{0}B}/{M}}\) to derive \DT~from an experimental value 
for $B$=450 kbar.   The \DT~value from neutron diffraction also represents a 
bulk measurement of lattice thermal behavior.\cite{Lawson94}  However, 
since it is derived from thermal displacement factors specifically attributed to the Pu 
atoms, it is referenced as $\Theta_{\textrm{DW}}$.  The \DT~values 
obtained from EXAFS and neutron resonant absorption\cite{Lynn98} are 
distinct from those derived from the other methods in that they are element specific, yet 
there are important differences to be noted.  EXAFS inherently 
measures only phonon modes that involve specific atom pairs (i.e., we must consider 
the reduced mass for the pair).  In contrast, neutron resonance absorption 
measures the average of all modes associated with a single atomic 
species (i.e., only the mass of the absorbing atom must be considered). \

\begin{table}
\caption{Comparison of Pu and Ga Specific Debye Temperatures Obtained 
for $\delta$-Pu by Various Techniques}
\begin{ruledtabular}
\begin{tabular}{lccl}
    $\delta$-Pu alloy & Pu-{$\Theta_{\textrm{D}}$} & 
    Ga-{$\Theta_{\textrm{D}}$} & Method \\\hline
    3.3 at\% Ga  &  110.7  &                 &  EXAFS  \\
                       &             &   202.6      &  EXAFS  \\
                       &   127    &                 &  Neut. 
Abs.\footnote{Neutron Resonance Doppler spectroscopy, Ref. \onlinecite{Lynn98}.}\\
                       &             &    255       &  Neut. Abs. \\
                       &   105    &                 &  Bulk 
Mod.\footnote{Using the relation from Ref. \onlinecite{Moruzzi88} and the experimental 
bulk modulus noted in the text.} \\
                       &   115    &                 &  
Elastic\footnote{Ultrasonic measurement of 3 principal elastic 
constants, $C_{ij}$, Ref. \onlinecite{Ledbetter76}.}    \\
    6.6 at\% Ga  &  127    &                 &     
Elastic\footnote{Derived from Young's modulus, $E$, and torsional 
modulus, $G$, Ref. \onlinecite{Taylor68}.}    \\   
    5.0 at\% Ga  &   132   &                 &   
Elastic\footnote{Also derived from $E$ and $G$, Ref. \onlinecite{Rosen69}.}  \\   
                        &   132   &                 &   Neut.
Diff.\footnote{Neutron diffraction, Ref. \onlinecite{Lawson94}.} \\
\end{tabular}
\end{ruledtabular}
\label{debye}
\end{table}

In general the Pu specific values of \DT~obtained from all the 
techniques are similar in that they are all quite low, ranging from 
104 to 132 K.  It is not our intent to discuss these differences since 
they come from such a wide variety of techniques and variability in 
samples.   The Ga-specific Debye temperatures do however require more careful 
discussion, in relation to the Pu-specific values.  As 
mentioned in Sec. I, neutron resonant absorption found a Ga specific 
\DT~value of 255 K $\pm$22 K which, compared to the 
$\sqrt{m_{\textrm{Pu}}/m_{\textrm{Ga}}}$ weighted value of 
236 K, does not exclude a slightly stiffer force field than the one around the Pu sites.  
Analogously, EXAFS yields a Ga specific $\Theta_{\textrm{cD}}$ of 
202.6 $\pm$3.7 K which may be compared to the expected value of 164 
K (i.e., the $\sqrt{\mu_{\textrm{Pu}}/\mu_{\textrm{Ga}}}$ weighted value).  The observation of a 
significantly higher $\Theta_{\textrm{cD}}$ for the Ga-Pu pair relative to that 
expected for a simple substitutional lattice model, coupled with the 
anomalously large lattice contraction around the Ga sites clearly 
suggest that the Ga sites reside in a significantly stronger force field. 
The shorter Ga-Pu bonds, in fact, inherently imply a strengthening of 
the Ga-Pu bonds relative to the Pu-Pu bonds.  \

Regarding factors that contribute to $\delta$-phase stabilization, 
these results are consistent with the Ga atoms having a significant influence 
on the local electronic structure.  One obvious possibility is that charge transfer 
exists between the Ga $pd$ states and the corresponding Pu $df$ 
states.  This orbital overlap will directly dictate 
the extent of lattice vibrational distortions and local geometry within the crystal 
structure, that is, with a preference towards a fcc lattice in the 
presence of Ga.  Therefore, the results of this EXAFS study give the first 
unambiguous evidence for the impact of the vibrational properties on 
the local lattice effects. \

From the results of this study, we may also estimate values for 
``pair-specific bulk moduli.''  Indeed, according to the empirical relation between bulk 
modulus and Debye temperature established by Moruzzi {\em{et 
al.}}\cite{Moruzzi88}, we deduce from our calculated Debye 
temperatures for Pu and Ga values of 498 and 1666 kbar, respectively, for the bulk modulus.
Since the Pu-Ga alloy we have been studying 
has a composition in the dilute limit, this implies that Pu is mostly 
surrounded by Pu atoms whereas each Ga site can be viewed as embedded 
in a Pu matrix.  In addition, the bulk modulus of pure Ga metal as 
determined by theory,\cite{Turchi01} 606 kbar, or experiment,\cite{Wyckoff62} 
613 kbar, is quite low in accordance with the low melting point of 308 K. 
Hence the dramatic difference between the two estimated bulk moduli 
provides an equivalent description for the increase in strength 
gained in going from Pu-Pu (or Ga-Ga bonds) to Ga-Pu bonds.  This 
unusually large increase in stability of a Pu (Ga) matrix by addition 
of Ga (Pu) is reflected in the existence of very stable Pu-Ga compounds 
(i.e., large heats of formation), some of them exhibiting 
congruent melting.  It would be interesting to extend this work to 
other alloy compositions to quantify more appropriately these 
findings and have a better understanding of the unusual synergistic 
effects due to alloying in Pu-Ga and related systems.

\section{Conclusion}

New pair specific Debye temperatures for the Ga-Pu and Pu-Pu pairs in 
$\delta$-Pu were determined using a correlated-Debye model fit to 
temperature dependent EXAFS Debye-Waller parameters.  Using this 
formalism, we obtain pair specific correlated-Debye 
temperatures, $\Theta_{\textrm{cD}}$, of 110.7 $\pm$1.7 K and 202.6 $\pm$3.7 K, for the 
Pu-Pu and Ga-Pu pairs, respectively.  The results for the 
$\Theta_{\textrm{cD}}$ Pu-Pu pair compare well with 
previous vibrational studies on $\delta$-Pu.  In addition, our results represent 
the first unambiguous determination of Ga-specific vibrational properties, 
i.e, Ga-Pu $\Theta_{\textrm{cD}}$ , in PuGa alloys.  Because the Debye temperature can be related 
to a measure of the lattice stiffness, these results indicate that the 
Ga-Pu bonds experience a stronger force field than the corresponding Pu-Pu 
bonds.  This effect has important implications for lattice stabilization 
mechanisms in these alloys.  If the vibrational properties in $\delta$-Pu are dependent on the 
type of impurity, further studies of Debye temperatures with other ``stabilizers'' 
will be important for discerning key aspects of the stabilization 
mechanism. \

\begin{acknowledgments}
This work was performed under the auspices of the U.S. Department of 
Energy (DOE) by the University of California Lawrence Livermore National Laboratory 
under contract No. W-7405-Eng-48.  This work was partially 
supported (C. H. B.) by the Office of Basic Energy Sciences, Chemical Sciences Division of the U. S. 
DOE, Contract No. DE-AC03-76SF00098.  The authors also wish to thank 
John Rehr and Alex Ankudinov for helpful discussions.  This work was done (partially) at SSRL, 
which is operated by the Department of Energy, Division of Chemical Sciences.
\end{acknowledgments}

\bibliographystyle{apsrev}

\end{document}